\documentclass[12pt]{article}
\usepackage{amssymb,amsmath}
\begin{document}

%%%%%%%%%%%%%%%%%%%%%%%%%%%%%%%%%%%%%%%%%%%%%%%%%%%%%%%%%%%%%%%%%%%%%
%%%%%%%%%%%%%%      TITLE PAGE     %%%%%%%%%%%%%%%%%%%%%%%%%%%%%%%%%%
%%%%%%%%%%%%%%%%%%%%%%%%%%%%%%%%%%%%%%%%%%%%%%%%%%%%%%%%%%%%%%%%%%%%%

\sloppy
\title
%{\hfill{\normalsize\sf FIAN/TD/01-15}    \\
 %           \vspace{1cm}
{\Large On the role of pressure in generating the gravitational field   }

\author
 {
       A.I.Nikishov
          \thanks
             {E-mail: nikishov@lpi.ru}
  \\
               {\small \phantom{uuu}}
  \\
           {\it {\small} I.E.Tamm Department of Theoretical Physics,}
  \\
               {\it {\small} P.N.Lebedev Physical Institute, Moscow, Russia}
  \\
  %       {\it {\small} 117924, Leninsky Prospect 53, Moscow, Russia.}
 }
%
%--------------------------------------------------------------------
\maketitle
%--------------------------------------------------------------------
%%%%%%%%%%%%%%%%%%%%%%%%%%%%%%%%%%%%%%%%%%%%%%%%%%%%%%
%ewcommand{\baselinestretch}{2.0}
%%%%%%%%%%%%%%%%%%%%%%%%%%%%%%%%%%%%%%%%%%%%%%%%%%%%%%%%%%%%%%%%%%%
%%%%%%%%%%%%%%%      TITLE PAGE     %%%%%%%%%%%%%%%%%%%%%%%%%%%%%%%%%%
%%%%%%%%%%%%%%%%%%%%%%%%%%%%%%%%%%%%%%%%%%%%%%%%%%%%%%%%%%%%%%%%%%%%%

%------------------------------------

%---------------------------------------------------
%%%%%%%%%%%%%%%%%%%%%%%%%%%%%%%%%%%%%%%%%%%%%%%%%%%%%%

%--------------------------------------------------------------------
%%%%%%%%%%%%%%%%%%%%%%%%%%%%%%%%%%%%%%%%%%%%%%%%%%%%%%
\begin{abstract}
The Einstein equations for static gravitational field depend on energy density 
and pressure. So one may expect that  solutions should depend on two parameters: mass and its analogue originated from pressure. Yet the  Schwarzschild solution have only mass parameter. So does its linear approximation. On the other hand the solutions of linearized  Einstein equations, obtained using graviton propagator, contain mass and its pressure analogue. This suggests that a phenomenological approach to gravity, using propagators and many graviton vertices, should lead to a theory different from general relativity.
\end{abstract}

%\section{Introduction }
 The integral of the energy density over space volume gives the mass of the body and this parameter is  incorporated in Schwarzschild solution. The second quantity- pressure is absent. Usually the role of pressure is negligible. But  when a star like our Sun is so compressed that its radius becomes of  order of its gravitational radius, the volume per nucleon becomes of order
$(\frac{\hbar}{mc})^3$ ( For this remark I am indebted to V.I. Ritus). So the particles of the body become relativistic. The pressure parameter $p$ becomes comparable with energy density $\epsilon$, see for example equation (35.9) in [1]. This pressure should influence the gravitational field not only inside the body, but also outside.
This is seen from  the linearized Einstein equations written  in the form ($g_{ik}=\eta_{ik}+h_{ik}$)
$$
\Delta h_{ik}=\frac{16\pi G}{c^4}\bar T_
{ik},\quad \bar T_{ik}=T_{ik}-\frac12\eta_{ik}T_l{}^l, \quad
\eta_{ik}=\rm diag(1,-1,-1,-1.),                                         \eqno(1)
$$
see for example, equations (95.8) and (105.11) in [1].
Here
$$
T_{ik}=\rm diag(\epsilon, p,p,p), \quad \bar T_{00}=\frac12(\epsilon+3p),\quad
\bar T_{11}=\bar T_{22}=\bar T_{33}=\frac12(\epsilon-p).                       \eqno(2)
$$

The solution of (1) is well known, 
$$
h_{ik}(\vec x)=-4\frac{G}{c^4}\int\frac{\bar T_{ik}(\vec x')}{|\vec x-\vec x'|}d^3x'. \eqno(3)
$$
It is similar to Coulomb problem. For spherically symmetric body it follows from (3) that outside the body
$$
h_{00}=-\frac{2Gm_t}{rc^2}, \quad h_{s}\equiv h_{11}=-\frac{2Gm_s}{rc^2}. \eqno(4)
$$
Here $m_t$ is the effective mass (with the pressure taken into account) defining $h_{00}$ and
similarly for space part of metric $h_s\equiv h_{11}$.
 When $p=0$ we have the linear approximation of Schwarzschild solution in isotropic coordinate system $h_{ik}=\delta_{ik}2\phi$, where $\phi$ is the Newtonian potential, see
eq.(100.3) in [1]. When $p\ne0$, $h_{00}\ne h_{11}$. The difference reflects the role of pressure in generating the gravitational field.

So, the presence of pressure $p$ of the order $\epsilon$ essentially affect the gravitational field. To go to the next approximation , we have to know  the 3-graviton vertex. It is difficult to believe that summing the contributions from all powers of gravitational constant will eliminate the influence of pressure in gravitational field outside the body. 

In any case one can think of a sphere filled with photon gas. The inner wall of the 
sphere can be made almost ideally reflecting the photons. For this gas $\epsilon=p/3$. 
For sufficiently large radius of the sphere one can make the $mc^2$ of the sphere
of order of energy of photon gas (for fixed photon density). The gravitational field 
of this ball may be weak everywhere and equations (1-4) are applicable. The influence of pressure  on gravitational field outside the body should be essential. But the linear approximation of the Schwarzschild solution says otherwise. This suggest that
the phenomenological approach to gravity, using only graviton propagators and many graviton vertices, will lead to a theory different from general relativity.

The remarks made for the Schwarzschild field concerns also Weyl's solution for an axially symmetric body [1-3]. We note that in contrast to Schwarzschild solution Weyl's solution has no
event horizon.
 If we doubt these solutions, 
it is not surprising that Weyl's solution do not goes over to Schwarzschild one
(in a natural way), even when the axial symmetry becomes indistinguishable from spherical one.

%\section{Conclusions}
 %\section*{Acknowledgments}
I am greatly indebted to V.I. Ritus for criticism, which helps me to formulate the discussed problem more clearly. The work was supported by Scientific Schools and Russian Fund for Fundamental Research (Grants 1615.2008.2 and 08-02-01118).

 \section*{References}
1. L.D.Landau and E.M.Lifshitz, {\sl The classical theory of
   fields}, Addison-Wesley, Cambridge, MA, 1971).\\
2.J.L. Synge, {\sl Relativity: The general theory,} North Holland Publishing Company,
   Amsterdam,1960\\
3. C. M\"oller, {\sl The theory of relativity}, Clarendon press, Oxford, 1972.\\

\end{document}